\documentclass[prd,twocolumn,showpacs,preprintnumbers,amsmath,amssymb]{revtex4}

\usepackage{graphicx}
\usepackage{dcolumn}
\usepackage{bm}

\catcode`\@=11
\def\lsim{\mathrel{\mathpalette\@versim<}}
\def\gsim{\mathrel{\mathpalette\@versim>}}
\def\@versim#1#2{\vcenter{\offinterlineskip
\ialign{$\m@th#1\hfil##\hfil$\crcr#2\crcr\sim\crcr } }}
\catcode`\@=12

\newcommand{\nn}{\nonumber}

\newcommand{\be}{\begin{equation}}
\newcommand{\ee}{\end{equation}}
\newcommand{\bea}{\begin{eqnarray}}
\newcommand{\eea}{\end{eqnarray}}

\newcommand{\Dsla}{D\hspace{-6pt}/~} 

\begin{document}
\input epsf.tex


\preprint{KANAZAWA-07-03}
\title{
Higgs and Top quark coupled with a conformal 
gauge sector}
\author{Haruhiko Terao}
\email{terao@hep.s.kanazawa-u.ac.jp}
\affiliation{
Institute for Theoretical Physics, Kanazawa
University, Kanazawa 920-1192, Japan
}%
\date{\today}
\begin{abstract}
We propose a dynamical scenario beyond
the standard model, in which the radiative
correction to the Higgs mass parameter
is suppressed due to a large anomalous dimension
induced through a conformal invariant coupling
with an extra gauge sector.
Then the anomalous dimension also suppresses
the Yukawa couplings of the Higgs field.
However, the large top Yukawa coupling can be 
generated effectively through mixing among top quarks 
and the fermions of the conformal gauge sector. 
This scenario is found to predict
a fairly heavy Higgs mass of about 500 GeV.
We present an explicit model and show consistency
with the Electro-Weak precision measurements of 
the S and T parameters as well as the Z boson decay width.

\end{abstract}



\maketitle

\section{Introduction}
Unnaturalness of the standard model (SM) 
indicates that at least the Higgs sector is
modified with new interactions and particles
above some energy scale $M$ not much higher than 1 TeV.
Especially, if the Higgs mass is less than $200$GeV, 
which is expected from the Electro-Weak (EW) precision tests
\cite{LEP},
then the scale $M$ should be no more than $O(1)$TeV.
The origin of fine-tuning is the quadratic divergence
in the radiative corrections to the Higgs mass parameter.
Therefore the new physics above the scale $M$ must
remove the quadratic divergence.
However absence of the quadratic divergence is not
sufficient phenomenologically.
The EW precision tests constrain some dimension 6 operators 
added effectively to the EW theory severely \cite{GW}
and the scale  of the new physics is expected to be higher
than $5 \sim 10$TeV generically \cite{LEPparadox}. 
The discrepancy between this scale and the scale $M$
for naturalness is called the LEP paradox or the
little hierarchy problem. 
Thus the new physics should have specific properties
in order to be consistent with the EW precision
tests simultaneously.

The quadratic divergence can be eliminated by imposing
global symmetries. 
The most postulated candidate would be supersymmetry,
which predicts a fairly light Higgs mass.
The little Higgs models \cite{littlehiggs} 
also assume global symmetries to suppress the quadratic 
divergence to mass of the Higgs boson 
appearing as a pseudo Nambu-Goldstone boson.
Then the Higgs mass is predicted to be light again, since  
the Higgs quartic coupling is also suppressed due to the
global symmetries. 
It should be also mentioned that the supersymmetric
extension does not remove the hierarchy problem completely.
The rigid supersymmetry restricts the Higgs mass to be
less than the Z-boson mass $M_Z$.
Meanwhile, the Higgs mass has been constrained to be heavier
than 115GeV by LEPII \cite{LEP}.
Therefore, the scale of the supersymmetry breaking parameters
must be rather large and a sizeable radiative correction
to the Higgs mass is induced.
Consequently, fine-tuning of a few percent is
required to realize the EW symmetry breaking of 250GeV.
There have been no convincing scenarios overcoming 
this problem, although various models have been 
proposed so far 
\cite{lowsusy,extradterm,fathiggs,supersoft,superlittle,KNT}.
So it would be worth while seeking for other possibilities 
as well.

Although the results of the EW precision tests are
consistent with the SM with a light Higgs boson,
the heavy Higgs boson is not always excluded. 
The recent analysis shows that 
a heavy Higgs mass of 400-600 GeV can be consistent 
\cite{heavyhiggs,BHR},
as long as there are extra contributions leading to 
$\Delta T = 0.25 \pm 0.1$ and a small $\Delta S$
\cite{LEP}.
Therefore the possibility to find a heavy Higgs at LHC
is still open, though the constraint to the new 
physics is rather restrictive.
{}For example, the Technicolor models, in which
the composite Higgs boson is rather heavy,
are known to suffer from too large 
corrections to the S-parameter \cite{PT,strongEW}.
Even if the Higgs boson is not a composite particle,
the Higgs mass as heavy as the triviality bound, which is 
about 600GeV, implies that a strong interaction
is involved with the Higgs sector at the TeV scale.
Then large oblique corrections are generated in general.
Meanwhile, the degree of fine-tuning to the Higgs
mass parameter is ameliorated for a heavy Higgs mass
and the scale of new physics may be raised up 
somewhat \cite{KM,BHR}. However the improvement is not
significant and new physics should appear at a few
TeV \cite{CEH}.

In this paper, we study a dynamical scenario to protect
the Higgs mass parameter from the quadratic 
diverging corrections.
Suppose that the Higgs field acquires a positive anomalous
dimension, then the degree of divergence of 
the mass correction is reduced.
Therefore, if the anomalous dimension is sufficiently
large, then cutoff dependence of the mass correction may be 
drastically suppressed and the Higgs
sector is relieved from the fine-tuning problem.
Indeed it will be shown that such a large anomalous dimension
can be realized by introducing a coupling of the Higgs with a 
strongly interacting conformal field theory (CFT).

Recently Luty and Okui \cite{LO}
has also discussed a Higgs model 
with a large anomalous dimension,
the ``conformal Technicolor", and it's 
AdS/CFT correspondence \cite{AdS/CFT}.
In their scenario, the Higgs boson is given as a fermion
composite.
In this paper, we consider a scenario in which 
the EW symmetry breaking (EWSB) is not dynamical
and the Higgs boson behaves as a point particle
even above the TeV scale.
Therefore the strongly coupled sector does not induce a large
correction to $S$-parameter.
In this respect, our scenario is distinct from the
models discussed in Ref.~\cite{LO} and
let us call it  ``the conformal Higgs model''
hereafter.
Explicitly, the CFT is assumed to be a strongly
coupled gauge theory with a appropriate number of 
vector-like fermions and the Higgs boson interacts through 
a large Yukawa coupling with some of these fermions.
Then it is found that the Higgs mass given in this
scenario is as heavy as 500GeV due to the strong 
interaction, although the cutoff dependence is suppressed
by the anomalous dimension.

In order to make such a scenario viable phenomenologically, 
we need to think about the following problems.
First, we note that the large anomalous 
dimension of the Higgs also suppresses the Yukawa
couplings with quarks and leptons. 
Therefore it seems that this scenario is incompatible
with the large top quark mass.
It may be suggestive that many approaches towards 
the little hierarchy problem are faced with difficulty
to explain the top quark mass simultaneously.
In order to avoid this problem, we suppose
that the origin of the prominently large top
quark mass is different from the Yukawa coupling with
the Higgs boson.
More explicitly, we consider that the dynamics
producing masses for the extra vector-like fermions
also induces mixing between top quarks and the
extra fermions \cite{KNT}.
Then the top Yukawa coupling
can be generated through the large Yukawa coupling of
the Higgs with the extra fermions effectively.

Another problem to be concerned is consistency with 
the EW precision tests.
As is mentioned above, the heavy Higgs boson requires
a suitable amount of extra contributions to the 
$T$-parameter.
In the present model, the custodial symmetry is
largely violated, since only top quark fields
are assumed to mix with the CFT sector significantly. 
Therefore the loop corrections with
the extra fermions contribute to the $T$-parameter.
The size of the correction depends on the masses
of these fermions, which gives also the decoupling
scale of the CFT sector from the  EW theory.
We will show an explicit model with the decoupling
scale of a few TeV can explain the $T$-parameter
consistent with the EW precision test.

Recently various alternative EWSB scenarios 
defined in the warped extra dimensions, or the
five-dimensional anti-de Sitter (AdS) space,
with four-dimensional boundaries \cite{AdS/CFT}
have been studied extensively,
{\it e.g.} 
the Higgsless models \cite{higgsless},
the minimal composite Higgs models \cite{minimalcomposite}.
The studies of these models also revealed that
it is a rather non-trivial problem to satisfy the
constraints by the precision measurements with 
realizing the large top quark mass simultaneously.

These models are also expected to have four dimensional
interpretation in terms of a strongly-coupled
CFTs according to the AdS/CFT correspondence. 
However explicit realizations of the
CFTs have not been known so far.
It is thought that the explicit CFT discussed in 
this paper offers an example of this class of models.
Although it would be also interesting to find the 
warped model related by the AdS/CFT correspondence 
conversely, we restrict our discussions within the 
four-dimensional model building in this paper.

This paper is organized as follows.
In section II, the IR fixed points in gauge-Yukawa theories 
are examined as an explicit mechanism to endow a
large anomalous dimension to the Higgs field.
There we apply the Wilson renormalization group (RG) 
equations in the ladder approximation to analyze
the fixed point.
The dynamical effects by the anomalous dimension
are discussed.
In section III, we present explicitly 
the conformal Higgs model, in which
the SM Higgs and the top quark are
coupled with a conformal invariant gauge theory 
through Yukawa interactions. 
We also introduce another strong interaction to break
down the conformal invariance at the TeV scale by
spontaneous mass generation.
The model is considered so that the proper mixings
between the extra fermions and top quarks are induced
with this mass generation.
In section IV, consistency with the EW precision 
measurements is examined by evaluating the loop corrections
by the extra fermions explicitly.
Finally section V is devoted to conclusions and discussions.

\section{Dynamics generating a large anomalous dimension}

\subsection{Power of divergence}
The reason why radiative corrections to the Higgs mass 
parameter $m^2_H$ show quadratic divergence is
that the dimension of the mass operator  is two.
Therefore, if the Higgs field carries a positive anomalous 
dimension $\gamma_H$, 
then this power of divergence is reduced.
Suppose that $\gamma_H$ is given to be a scale independent
constant, then the correction to $m^2_H$, $\delta m^2_H$,
depends on the cutoff scale $\Lambda$ as
\be
\delta m^2_H \sim \left(
\frac{\Lambda}{\mu}
\right)^{\epsilon}~\mu^2,
\ee
where the power of divergence is given by 
$\epsilon = 2(1 - \gamma_H)$
\footnote{We defined the anomalous dimension $\gamma_H$ 
as $\gamma_H = - (1/2) d \ln Z_H/ d \ln \mu$ and
the dimension of $H^2$ is given by $2(1 + \gamma_H)$.},
and $\mu$ denotes the renormalization scale.
Note that this correction is suppressed and
may be approximated as a logarithmic
one, when the anomalous dimension is close to 1.

Thus, the cutoff scale $\Lambda$ can be raised up without
severe fine-tuning of the parameters with help of the
anomalous dimension \cite{LO}.
One may wonder whether a concrete dynamics realizing such a 
large anomalous dimension exists or not.
The coupling constants inducing such an anomalous 
dimension must
be not only fairly large but also scale independent.
This implies that we should consider conformal field
theories, in which the coupling constant of the Higgs field
is stabilized at an infrared (IR) attractive fixed point.
In this section, we examine an explicit example of the 
gauge-Yukawa theory.

\subsection{IR fixed point with non-trivial Yukawa
coupling}

It has been known for some time that an IR attractive fixed
point exists for the QCD like theory with an appropriate
number of flavors, which is called the Banks-Zaks (BZ)
fixed point \cite{BZ}. 
Then the gauge theory becomes a CFT at low energy 
irrespectively of the coupling given at high energy.
For the $SU(N_c)$ gauge theory with $N_f$ Dirac fermions $\psi$
of the fundamental representation, the two-loop beta function
shows that the fixed point exists for the number of flavors 
$N_f$ within $(34/13) N_c < N_f < (11/2) N_c$
in the large $N_c$ leading.
Besides the recent studies by numerical simulations of the 
$SU(3)$ lattice gauge theories also indicates that there is 
the non-trivial fixed point for $7 \leq N_f \leq 16$
\cite{lattice}.

At the fixed point, the fermion mass operator
$\bar{\psi}\psi$ acquires a negative anomalous
dimension through the gauge interaction. 
Therefore the perturbation by a Yukawa operator
$\lambda \bar{\psi} \psi \phi$ with
a singlet scalar $\phi$ is relevant there.
The Yukawa coupling $\lambda$ is enhanced 
rapidly towards the IR direction.
Since this Yukawa coupling induces a positive anomalous dimension
to the scalar field, it may be expected that 
the flow of $\lambda$ eventually approaches another
fixed point $\lambda_*$, which is  IR attractive.

This new fixed point can be explicitly shown,
when the gauge theory has the BZ 
fixed point in the perturbative region.
Let us consider the following Yukawa interaction
with a gauge singlet complex scalar $\phi$,
\be
{\cal L} \sim 
- \phi \sum_{i=1}^{n_f} \lambda_i ~\bar{\psi}_{Li} \psi_R^i 
+ \mbox{h.c.},
\label{singletyukawa}
\ee
where $n_f = 1, \cdots, N_f$.
When we evaluate the beta function for the Yukawa coupling
$\alpha_{\lambda_i}=|\lambda_i|^2/(4 \pi)^2$ 
in the one-loop approximation 
and the beta function for the gauge coupling
$\alpha_g = g^2/(4 \pi)^2$ in 
the two-loop approximation, 
then they are found to be
\bea
\mu \frac{d \alpha_g}{d \mu} 
&=&
-2 b_0 \alpha_g^2 - 2 b_1 \alpha_g^3 
- 2  \alpha_g^2 \sum_{i=1}^{n_f} 
\alpha_{\lambda_i}, 
\label{pgaugebeta}
\\
\mu \frac{d \alpha_{\lambda_i}}{d \mu}
&=&
2 \alpha_{\lambda_i} \left[
2 N_c \sum_{j=1}^{n_f} \alpha_{\lambda_j} + \alpha_{\lambda_i}
- 6 C_2(N_c) \alpha_g
\right],
\label{pyukawabeta}
\eea
where the coefficients are given with the
quadratic Casimir $C_2(N_c) = (N_c^2 -1)/2N_c$ as
\bea
b_0 &=& \frac{11}{3} N_c - \frac{2}{3}N_f,  \\
b_1 &=& \frac{34}{3} N_c^2 - \frac{N_f}{2}
\left(
4 C_2(N_c) + \frac{20}{3} N_c
\right).
\eea
The BZ fixed point ($\lambda_i=0)$ 
exists when $b_0 >0$ and $b_1 < 0$.
The above  beta functions satisfy 
the fixed point given by
\bea
& &
\alpha_g^* = \frac{b_0}{-b_1 + \frac{6C_2(N_c) n_f}{2N_c n_f + 1}}
\simeq \frac{b_0}{-b_1 + 3/2}, 
\label{pgaugefp}\\
& &
\alpha_{\lambda_i^*} = \alpha_{\lambda^*} 
= \frac{6C_2(N_c)}{2 N_c n_f + 1} \alpha_g^*
\simeq \frac{3}{2n_f}\alpha_g^*,
\label{pyukawafp}
\eea
where the last expressions stand for the large $N_c$
leading part.
It is also easily shown that
this fixed point is IR attractive.
The anomalous dimension of the scalar filed at the
IR fixed point is given by 
$\gamma_{\phi}^* = 2 N_c n_f \alpha_{\lambda^*}$.
Hereafter we restrict the  Yukawa couplings to be
the same, $\lambda_i = \lambda$, since
they are identical at the IR fixed point.

\subsection{Non-perturbative evaluation of the anomalous 
dimension by the RG method}

We are interested in the anomalous dimension 
$\gamma_{\phi}$ obtained in the
strongly coupled region, where the perturbative analysis
is not valid.
However it is a quite difficult problem to extend the 
RG equations to ones fully reliable even in the 
non-perturbative region.
In practice, the non-perturbative dynamics of chiral symmetry
braking phenomena in the QCD with many flavors has been 
studied so far by solving the Dyson-Schwinger equations
mostly \cite{ATW,MY}.
Unfortunately, dynamics around the fixed point cannot be
captured by the DS equations.
This is because the DS equations are given
with respect to the order parameter, which is
vanishing around  the IR fixed point.

It has been found \cite{NPRG,3dQED,Gies} that the
Exact Renormalization Group (ERG) \cite{ERG},
which offers an explicit formulation of the Wilson RG,
is also applicable to study of the chiral
symmetry breaking phenomena.
Besides, the phase structure as well as the order
parameters obtained by solving the RG equations and the
DS equations are found to be identical
within the so-called (improved) ladder approximation.
Moreover, the RG equations enable us to study renormalization 
properties directly irrespectively of the phases.
Therefore the RG approach has a great advantage
to examine dynamics around the fixed point \cite{3dQED}.

The ERG equation gives evolution of the Wilsonian
effective action under infinitesimal shift of the cutoff
scale by a functional form.
It is necessary to reduce the equations by some
approximation in the practical analysis.
It is usually performed to truncate the series of
local operators in the Wilsonian effective action.
Then improvement of the approximation is made by 
increasing the level of the operator truncation.
Once the operator truncation is performed,
then the ERG equation turns out to be a set of
one-loop RG equations.
Difference from the perturbative RG lies in
that couplings of the higher dimensional operators
are involved as well.
This enables us to sum up an infinite number of
loop diagrams.

It was found through the previous studies \cite{NPRG,3dQED}
that the effective four-fermi operators are found to 
play an important role for the non-perturbative 
analysis of the chiral symmetry breaking.
The reason may be understood by thinking over
the anomalous dimension $\gamma_{\bar{\psi}\psi}$.
Fig.~1 shows schematically 
how the anomalous dimension 
is represented in terms of the effective four-fermi
couplings in the NPRG framework.
The four-fermi couplings are also given as a sum of
infinitely many loop diagrams by solving the RG
equations.
Thus a non-perturbative sum of the loop diagrams
is carried out by incorporating the four-fermi 
operators.

\begin{figure}[htb]
\includegraphics[width=0.4\textwidth]{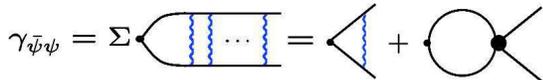}
\caption{\label{fig1} 
The anomalous dimension $\gamma_{\bar{\psi}\psi}$
in terms of the effective four-fermi vertex is
shown schematically.
}
\end{figure}

First we consider the Wilsonian effective 
action of the $SU(N_c)$
gauge theory with $N_f$ massless flavors.
The induced operators in the 
effective action should be invariant under the
color $SU(N_c)$ as well as the
flavor symmetry $U(N_f)_L \times U(N_f)_R$.
Even if we restrict them to the four-fermi interactions,
there are various invariant operators.
In order to perform the minimal analysis,
we take the following effective Lagrangian,
\bea
{\cal L}_{\rm eff} 
&=&
-\frac{1}{4g^2} {\rm tr} F_{\mu \nu} F^{\mu \nu}
+ i \sum_i \bar{\psi}_i \Dsla \psi^i \nn \\
& &
- \frac{2G_S(\Lambda)}{\Lambda^2} 
\sum_{i,j} 
\bar{\psi}_{Li} \psi^j_R \bar{\psi}_{Rj} \psi^i_L 
+ \cdots,
\label{effaction}
\eea
where $i, j = 1, \cdots, N_f$ denote the flavor
indices.
In practice, the RG analysis
beyond this simple approximation can be also
performed. 
A detailed study of QCD like theories and the
gauge-Yukawa theories with many flavors by the
NPRG method is reported separately \cite{TT}.
There it is found to be sufficient to incorporate
the above four-fermi operator and to sum up the
ladder diagrams to examine the IR fixed point.

We should also incorporate higher dimensional
operators of the field strength, such as
$(D_{\mu} F^{\mu \nu})^2$, 
$(F_{\mu \nu}F^{\mu \nu})^2$ and so on
in order to extend the gauge beta function beyond
the one-loop level.
However practical calculations 
are rather tedious and face with the problem in
maintaining the gauge (or BRS) invariance.
Therefore we do not deal with the ERG equations
faithfully, but substitute the two-loop beta
function given by (\ref{pgaugebeta})
for the RG equation of the gauge coupling instead.

The explicit RG equation for the four-fermi coupling
$G_S/4\pi^2$ is easily found by calculating
one-loop diagrams and is given simply by \cite{NPRG}
\be
\Lambda \frac{d g_S}{d \Lambda}
= 2g_S
- 2 N_c \left(
g_S + \frac{3}{2} \alpha_g
\right)^2,
\label{ladderRG}
\ee
where we used the Landau gauge propagator.
This beta function leads to the fixed points of 
$g_S$ as
\be
g_S^* = \frac{1}{2N_c}\left(
1 - \frac{\alpha_g^*}{2 \alpha_{g,{\rm max}}^*}
\pm \sqrt{1- \frac{\alpha_g^*}{\alpha_{g,{\rm max}}^*}}
\right),
\label{gsfp}
\ee
where $\alpha_g^*$ denotes the gauge coupling
$g^2/(4\pi)^2$ at the fixed point, which is
determined by the two-loop beta function
(\ref{pgaugebeta}).
The solution with $+(-)$ sign gives the UV (IR)
fixed point value.
The number represented by $\alpha_{g,{\rm max}}^*$
in (\ref{gsfp}) 
is the maximal value of the gauge coupling of the
fixed point, which is given by
\be
\alpha_{g,{\rm max}}^* \equiv 
\frac{g_{\rm max}^{*2}}{(4\pi)^2}
=\frac{1}{12C_2(N_c)}.
\ee

In Fig.~2, the RG flows of $(\alpha_g, N_c g_S)$
obtained by solving RG equations (\ref{pgaugebeta})
and (\ref{ladderRG})
are shown in the case of  $N_c=3$ and $N_f=12$. 
The points A, B and C stand for the non-trivial
fixed points of the RG equations.
It is seen that the phase boundary appears 
and all flows in the lower phase 
approach the IR fixed point A which is the BZ fixed
point.
It is found that the lower (upper) area of the phase
boundary is unbroken (broken) phase of the chiral
symmetry \cite{NPRG}.
The phase boundary also shows that the critical gauge 
coupling of the cutoff gauge theory is given by 
$\alpha_g^{\rm cr} = (g^{\rm cr})^2/(4\pi)^2 \sim 0.09$.

\begin{figure}[htb]
\includegraphics[width=0.4\textwidth]{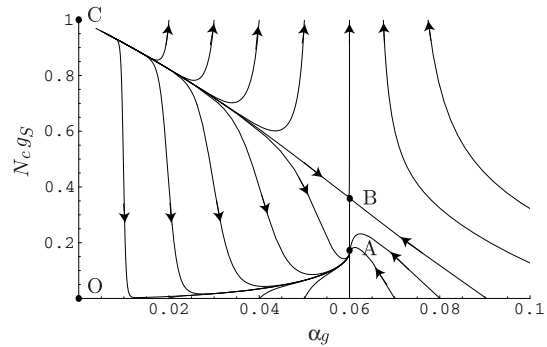}
\caption{\label{fig2}
The RG flows of the couplings 
$(\alpha_g, g_S)$ are shown.
The IR and the UV fixed points are represented 
by blob points A and B respectively.
The point C stands for the fixed point of the
four-fermi theory.
}
\end{figure}

Aspect of the RG flows varies with the flavor
number $N_f$.
When the gauge coupling of the fixed point of the
gauge beta function (\ref{pgaugebeta}) exceeds
$\alpha_{g,{\rm max}}^*$, then the
unbroken phase disappears.
Thus the simple RG analysis given above
leads to the conformal window of
\be
\frac{100N_c^2 -66}{25N_c^2-15} N_c < N_f < \frac{11}{3} N_c,
\ee
which coincides with the result obtained from the
DS equation in the improved ladder approximation
using the two-loop gauge beta function \cite{ATW}.
However we should note also that the lower bound 
is dependent on the gauge beta function considerably
\footnote{When we apply the three-loop gauge beta function,
then the bound of the conformal window is 
fairly reduced.
Although the three-loop result is not necessarily more 
reliable.}.

In the non-perturbative RG framework, the anomalous 
dimension of the fermion mass operator
$\gamma_{\bar{\psi}\psi}$ is given with the four-fermi
coupling as
\be
\gamma_{\bar{\psi}\psi}
= -6C_2(N_c) \alpha_g - 2N_c g_S. 
\ee
Therefore the explicit value at the BZ fixed 
point is found to be
\be
\gamma_{\bar{\psi}\psi}^*
= -1 + \sqrt{1-\frac{\alpha_g^*}{\alpha_{g,{\rm max}}}},
\ee
which shows that $-1 < \gamma_{\bar{\psi}\psi}^* < 0$
in the conformal window \cite{DSanomalous}.
In the case of $N_c=3$ and $N_f = 12$, 
$\gamma_{\bar{\psi}\psi}^* = -0.8$ 
$(\alpha_g^* = 0.06 < \alpha_{g,{\rm max}}^* = 0.0625)$, 
which is fairly close to the critical value.

Now we consider to incorporate the Yukawa interaction
given by (\ref{singletyukawa}) to the above analysis.
Then the scalar exchange diagrams also induce four-fermi
operators. However the operators do not contain
$\bar{\psi}_{Li} \psi^j_R \bar{\psi}_{Rj} \psi^i_L$,
but truncated ones such as 
$\bar{\psi}_{Li} \gamma_{\mu} \psi^i_L 
\bar{\psi}_{Rj} \gamma^{\mu} \psi^j_R$.
Thus  the RG equation for $g_S$ (\ref{ladderRG}) 
is found to be intact even with the Yukawa interaction
except for the anomalous dimension of the fermion
$\gamma_{\psi} = \alpha_{\lambda}/2$;
\be
\Lambda \frac{d g_S}{d \Lambda}
= (2 + 2 \alpha_{\lambda}) g_S
- 2 N_c \left(
g_S + \frac{3}{2} \alpha_g
\right)^2.
\label{ladderRG2}
\ee
We may refer Ref.~\cite{TT} for the detailed analysis.
On the other hand,
the RG equation for the Yukawa coupling is modified
with the four-fermi interaction as
\be
\Lambda \frac{d \alpha_{\lambda}}{d \Lambda}
= 2 \alpha_{\lambda}(\gamma_{\phi} + \gamma_{\bar{\psi}\psi}),
\label{yukawaRG}
\ee
where the anomalous dimensions are given explicitly by
\bea
\gamma_{\phi} & = &
 2 N_c n_f \alpha_{\lambda}, 
\label{gammaphi}\\
\gamma_{\psi} &=& 
- 6 C_2(R) \alpha_g  +  \alpha_{\lambda} - 2 N_c g_S.
\label{gamma2psi}
\eea
Therefore we may solve the RG equations (\ref{pgaugebeta}),
(\ref{ladderRG2}) and (\ref{yukawaRG}).

In Fig.~3, the RG flows in the coupling space of
$(\alpha_g, \alpha_{\lambda}, N_c g_S)$
are shown in the case of $N_c=3, N_f=n_f=12$.
The black points stand for the fixed points.
It is seen that the IR fixed point A' exists 
and the gauge coupling takes almost the same 
value as that of the BZ fixed point A.
The points B and B' represent the UV fixed point
and the renormalized trajectories linking these
fixed points are also shown in Fig.~3.

\begin{figure}[htb]
\includegraphics[width=0.4\textwidth]{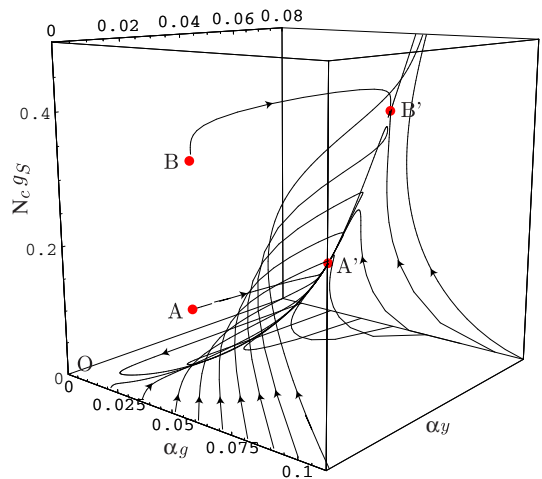}
\caption{\label{fig3}
The RG flows of in the coupling space of
$(\alpha_g, \alpha_{\lambda}, N_c g_S)$
are shown. 
The points A' and B' represent the IR and the UV 
fixed points of the gauge-Yukawa theory respectively,
while A and B stand for the BZ (IR) and UV
fixed points shown in Fig.~2.
}
\end{figure}

The anomalous dimensions at the IR fixed point
may be evaluated as
$ \gamma_{\phi}^* = - \gamma_{\bar{\psi}\psi}^* 
\simeq 0.8$ ($\epsilon = 2(1-\gamma_{\phi}^*) \simeq
0.4$).
It is noted that the Yukawa coupling is not
extremely large and $\lambda^* \simeq 1.3$.

\subsection{Effects of the large anomalous dimension}

Around the IR fixed point, the scalar potential is
renormalized in a peculiar way due to the large 
anomalous dimension \cite{TT}.
If the scalar potential at the scale $\lambda$
is given as
\be
V(\phi) =
\tilde{m}_{\phi}^2 \Lambda^2
|\phi|^2 + \frac{\lambda_4}{4} |\phi|^4
+ \cdots,
\ee
then the RG equations for the dimensionless 
parameters $\tilde{m}_{\phi}^2$ and $\lambda_4$
are found to be
\bea
\Lambda \frac{d \tilde{m}^2_{\phi}}{d \Lambda} &=&
-2(1 - \gamma_{\phi}) \tilde{m}^2_{\phi} + 
4N_c n_f \alpha_{\lambda} - 2 \tilde{\lambda}_4, 
\label{scalarmassRG}\\
\Lambda \frac{d \tilde{\lambda}_4}{d \Lambda} &=&
4 \gamma_{\phi}\tilde{\lambda}_4 
- 8 N_c n_f \alpha_{\lambda}^2 + \tilde{\lambda}_4^2,
\label{quarticRG}
\eea
where $\tilde{\lambda}_4 = \lambda_4/(4\pi)^2$ and
$\gamma_{\phi} = 2 N_c n_f \alpha_{\lambda}$.

For the simplicity, we shall discuss the solutions
of these equations in the large $N_c$ limit, or
with neglecting corrections by the quartic coupling
$\lambda_4$.
If we set the Yukawa coupling to the fixed point
value, then these equations are easily solved. 
The solution of the scalar mass at a low energy scale
$M$ is given with the initial parameter
$\tilde{m}^{2}_{\phi}(\Lambda)$
at the cutoff scale $\Lambda$
and may be written down as 
\be
m_{\phi}^2(M) = \left[
\tilde{m}^{*2}_{\phi} + 
\left( \frac{\Lambda}{M} \right)^{\epsilon}
(\tilde{m}^{2}_{\phi}(\Lambda) - \tilde{m}^{*2}_{\phi})
\right]M^2, 
\label{scalarmass}
\ee
where 
$\tilde{m}_{\phi}^{*2} \equiv 2 \gamma_{\phi}^*/\epsilon
= \gamma_{\phi}^*/(1 - \gamma_{\phi}^*)$.
It is explicitly seen that the power of divergence
is reduced to $\epsilon$ due to the anomalous dimension.
As the renormalization scale $\mu$ goes to zero,
the scalar mass 
$m_{\phi}^2(\mu) = \tilde{m}_{\phi}^{*2}\mu^2$ 
also goes to zero irrespectively of the initial value 
$\tilde{m}_{\phi}^2(\Lambda)$.

In the next section, we consider a model in which the
conformal invariance is terminated at a low energy
scale $M$ by adding masses to the fermions.
Then the scalar (mass)$^2$ may be estimated as
$O(1 - 10) \times M^2$.
The dependence on the cutoff scale $\Lambda$
as well as the initial parameter is remarkably weaken
for a small $\epsilon$.
To be explicit, 
$(\Lambda/M)^{\epsilon} \sim 6.3 (2.5)$ for
$(\Lambda/M)^2 = 10^4 (10^2)$ in the case of
$\epsilon=0.4$ ($N_c =3, N_f=12$).

On the other hand the anomalous dimension 
makes the quartic coupling 
highly irrelevant, since the scaling dimension
of $\lambda_4$ is given by $4 \gamma_{\phi}^*$
at the IR fixed point.
This does not mean that the quartic coupling
is eliminated.
The RG equation (\ref{quarticRG}) tells us that
$\lambda_4$ approaches the fixed point
value given by 
\be
\lambda_4^* \simeq
|\lambda^*|^2
= \frac{(4\pi)^2}{2N_c n_f }\gamma_{\phi}^*,
\ee
very strongly. 
Thus it is found that the value of the quartic
coupling $\lambda_4$ appearing below the scale $M$ is 
large in general.
Similarly the couplings of higher
point interactions of the scalar converge to
the fixed point values strongly.
The presence of the quartic coupling in
the RG equations (\ref{scalarmassRG}) and (\ref{quarticRG}) 
does not alter the above properties.

Now we shall move on to the extension of the SM.
We introduce a strongly coupled gauge sector in the
conformal window other than the SM and also 
assume that the Higgs field has a Yukawa coupling
with this sector just like (\ref{singletyukawa}).
In order to avoid extreme fine-tuning, scale of
the fermion mass $M$, which is the scale of
conformal symmetry breaking, should be of 
O(1)TeV. 
On the other hand, a large quartic coupling is 
induced through this Yukawa interaction.
Therefore, the observed Higgs mass should be as
large as the triviality bound with the
cutoff scale $M$.
For $M$ of a few TeV, the Higgs mass is expected
to be as heavy as 500 GeV \cite{KM}.
Then, the corrections to the Higgs mass
through the standard model interactions
does not require fine-tuning.

One may wonder if the extra fermions coupled 
with Higgs field induces so large corrections
that contradict with the EW precision
measurements.
On the other hand, the extra sector should generate
a suitable amount  corrections to the
$S$ and $T$ parameters, as was mentioned in Introduction.
We will discuss this phenomenological issue
in section~IV.

Another problem to be concidered is the mass of Top 
quark.
It is noted 
that the large anomalous dimension of the Higgs
field also suppresses the top Yukawa coupling,
which should be about 1 at low energy 
in order to explain the observed mass.
To be explicit, the anomalous dimension of the
Higgs $\gamma_H^*$ suppresses the top Yukawa coupling
as
\be
y_t(\mu) \sim \left(
\frac{\mu}{\Lambda}
\right)^{\gamma_H^*}
y_t(\Lambda).
\ee
Then, the top Yukawa coupling must be non-perturbatively
large immediately above the TeV scale.
Therefore we have to replace
the strongly coupled conformal sector so as to 
include the top quarks.
In general, scenarios of the EWSB with the Higgs 
field coupled to the strongly interacting sector, 
such as 
the walking Technicolor models \cite{walkingTC}, 
the conformal Technicolor models \cite{LO}
and so on, seem to face with a similar problem in 
compatibility with the top quark mass
\footnote{
The top quark must be treated differently from
other quarks in 
the models in the warped extra dimensions such as 
the higgsless models \cite{higgsless} and
the composite Higgs models \cite{minimalcomposite}.
Some supersymmetric models aiming to solve the
supersymmetric little hierarchy problem also possess
similar problem \cite{fathiggs,KNT}.
}.

In the next section, we consider a model, in which
the top Yukawa coupling is generated through
mixing between top quarks and the fermions in the
CFT sector coupled strongly with the Higgs \cite{KNT}.
On the way around, such a mixing effect
may explain origin of the prominently
large top quark mass.

\section{A phenomenological model}

Now we shall consider an explicit phenomenological
realization of the mechanism discussed in 
the previous section, namely the
conformal Higgs model.
We assume a strongly coupled gauge theory with
the gauge group of $SU(3)_{\rm CFT}$ as the
conformal sector.
The Higgs field couples with the vector-like
fermions of the CFT and behaves as a point-like
particle below a certain high energy scale $\Lambda$.
Here we do not think about the ultraviolet completion
above the cutoff scale $\Lambda$, though
the Higgs field may be generated as a fermion
composite at this scale.
The basic feature of the scenario is that
the EW symmetry is not broken by the strong
dynamics, but by the vacuum expectation value
(VEV) of the point-like Higgs.
The dynamics of the CFT sector plays a role in
suppressing the mass scale of the Higgs boson.

The color gauge group of the SM, $SU(3)_C$, 
emerges after spontaneous symmetry breaking (SSB)
of  $SU(3)_{\rm CFT} \otimes SU(3)'_C$ at the
TeV scale.
The quark fields are charged under $SU(3)'_C$,
while the vector-like fermions are charged
under $SU(3)_{\rm CFT}$.
These fermions can mix with each other below
the scale of SSB, since  both fermions are charged
under the color gauge group  $SU(3)_C$.
Then a large top Yukawa coupling is induced 
through the mass mixing between quarks and 
the colored heavy fermions from the CFT sector.

Explicitly, we will discuss the case in which mixing between
the weak doublets, $\Phi_L$ and $Q_{3L}$, and between the
weak singlet $\phi_R$ and $u_{3R}$ occurs dynamically.
One may suppose that similar mixing among all flavors 
of quarks and leptons also occurs and, moreover, the 
differences in these mixings explain the mass hierarchy.
However the mixing in the other quarks than the
top quarks should be small anyway, and we neglect them 
completely in this paper.

We also introduce another gauge interaction of
the group $G_{\rm DSB}$, which becomes strong at 
the TeV scale and breaks down the conformal 
symmetry by through fermion condensations.
The dynamics of the CFT is not influenced above
the TeV scale, since the gauge coupling of
$G_{\rm DSB}$ sector is rather weak there.
However, this dynamics induces masses of the 
extra fermions of the CFT sector and mixing 
with the quarks simultaneously at the TeV scale.

We assume that $G_{\rm DSB}$ is also an $SU(3)$
group and introduce the following vector-like 
fermions charged as,
\be
\begin{array}{c|ccccc}
& SU(3)_{\rm DSB} & SU(3)_{\rm CFT} & SU(3)_C' & SU(2)_W 
& U(1)_Y  \\ \hline
\psi^A  & \mbox{\bf 3} & \mbox{\bf 3} & \mbox{\bf 1} 
& \mbox{\bf 1} & \mbox{any}  \\ 
\eta^a  & \mbox{\bf 3} & \mbox{\bf 1} & \mbox{\bf 3} 
& \mbox{\bf 1} & \mbox{any} \\
\hline 
\Phi^{iA} & \mbox{\bf 1} & \mbox{\bf 3} & \mbox{\bf 1} & 
\mbox{\bf 2} & 1/6  \\ 
\phi^{iA} & \mbox{\bf 1} & \mbox{\bf 3} & \mbox{\bf 1} & 
{\bf 1} & 2/3 
\end{array}
\label{extrafermions}
\ee
where $A=1,2,3$ and $a=1,2,3$ stand for indices 
of $SU(3)_{\rm CFT}$ and $SU(3)_C'$ respectively.
Another suffix $i$ runs $1,2,3$.
Then the $SU(3)_{\rm CFT}$ gauge sector may be
regarded as  a gauge theory with
$N_c=3$ and $N_f=12$.
Therefore an IR fixed point appears at the fairly
strong coupling region.
The assignment of the $U(1)_Y$ hypercharges
for $\Phi$ and $\phi$ allows 
mixing with the up-sector quarks.

The Higgs field is allowed to have the following
Yukawa interactions,
\bea
-{\cal L}_{\rm Yukawa} 
&=&
Y_t \bar{Q}_{3L} u_{3R} \tilde{H} 
+ Y_b \bar{Q}_{3L} d_{3R} H \nn \\
& &
+ \lambda \sum_i \bar{\Phi}_{iA} \phi^{iA} \tilde{H},
\eea
where $\tilde{H} = \epsilon_{ab}H^{b*}$ as in the SM 
Lagrangian and we omitted quarks in the first and second 
generations.
For the sake of simplicity, we suppose the coupling 
$\lambda$ to be the IR fixed point value 
$\lambda_*$ at the scale of $\Lambda$ already.
The explicit value of $\lambda_*$ may be estimated
from Eq.~(\ref{yukawaRG}) by noting 
the number of flavors coupled with the Higgs is 
$n_f=3$, 
and found to be about $2.65$.
We will also neglect the Yukawa couplings 
of the third generation
$Y_t$ and $Y_b$ for a while, since these couplings are
suppressed by the anomalous dimension of the Higgs field.

The gauge interaction of $SU(3)_{\rm DSB}$
induces condensation of fermion bilinears
such as 
$\langle \bar{\psi}_A \cdot \psi^A \rangle$
and $\langle \bar{\eta}_a \cdot \eta^a \rangle$
at low energy,
where $\cdot$ represents $SU(3)_{\rm DSB}$-singlet
contraction.
The symmetry breaking of 
$SU(3)_{\rm CFT} \otimes SU(3)_C'$ to
$SU(3)_C$ also takes place, if the diagonal components of the
composite fields acquire non-vanishing 
VEVs as
\be
\langle  \bar{\psi}_A \cdot \eta^a \rangle
\equiv \langle \Omega_A^a \rangle \propto \delta_A^a, ~
\langle \bar{\eta}_a \cdot \psi^A \rangle
\equiv \langle \bar{\Omega}^A_a \rangle \propto \delta_a^A.
\ee
On the other hand, the effective
Lagrangian given at the scale
$\Lambda$  may contain the non-renormalizable interactions
such as
\bea
&-& 
{\cal L}_{\rm four-fermi} \nn \\
&=&
 \frac{c}{\Lambda^2} \bar{Q}_{3La} \Phi^{3A}_R 
(\bar{\psi}_A \cdot \eta^a)
+ \frac{\bar{c}}{\Lambda^2} \bar{\phi}_{3A} u_{R}^{3a} 
(\bar{\eta}_a \cdot \psi^A) \nn \\
& &
+ \frac{c_{\Phi}}{\Lambda^2} \bar{\Phi}_{iA} \Phi^{iA}
(\bar{\eta}_a \cdot \eta^a) 
+ \frac{c_{\phi}}{\Lambda^2} \bar{\phi}_{iA} \phi^{iA}
(\bar{\eta}_a \cdot \eta^a).
\label{4fermi}
\eea
The coefficients of the four-fermi interactions
are unknown, unless
the ultraviolet completion of the model at the scale
$\Lambda$ is fixed.
In any case, their explicit
values are unimportant in the following argument.
However, we assumed here that only the third generation
quarks have four-fermi interactions with
sizable couplings with some reason.
Otherwise large Yukawa couplings are induced to all flavors
as well as top quark through mixing effect discussed below.

Thus the low energy effective interactions 
obtained after the symmetry breaking may be reduced
to the form of
\bea
- {\cal L}_{\rm int} &=&
\lambda_* (\bar{\Phi}_{iL} \phi_R^i + \bar{\Phi}_{iR} \phi_L^i)
\tilde{H} + V(H) \nn \\
& &
+M_{\Phi} \left(
\bar{\Phi}_{3L} + \frac{ \omega}{M_{\Phi}} \bar{Q}_{3L}\right) 
\Phi_{R}^3 \nn \\
& &
+M_{\phi} \bar{\phi}_{3L} \left(
\phi_R^3 + \frac{\bar{\omega}}{M_{\phi}} u_{R}^3 \right),
\label{effint}
\eea
where we defined the parameters $\omega$ and 
$\bar{\omega}$ by the VEV of SSB as
$c\langle \Omega_A^a \rangle/\Lambda^2 = \omega \delta_A^a$
and
$\bar{c} \langle \bar{\Omega}^A_a \rangle/\Lambda^2 =
\bar{\omega} \delta_a^A$.
The mass parameters $M_{\Phi}$ and $M_{\phi}$
are also generated through the fermion condensation
$\langle \bar{\eta}_a \cdot \eta^a  \rangle$
and give the decoupling scale of the extra sectors
from the SM.
Naively these mass parameters are supposed to be
of the same order as $\omega$ and $\bar{\omega}$.
In the above argument, we did not take account of
the fermion condensation of 
$\langle \bar{\psi}_A \cdot \psi^A  \rangle$
intentionally,
since there are subtle dynamical issues.
We shall discuss the issues in the end of this section.

Now it is apparent from the effective Lagrangian 
(\ref{effint}) that mixing occurs between the quark fields
and the extra matter fields.
Note that this mixing respects the EW gauge symmetry,
since the SSB does not break the EW symmetry.
The mixing appears between the left-handed doublets, 
$Q_{3L}3$ and $\Phi_{3L}$, and also between the 
'Æright-handed singlets, $u_{3R}$ and $\phi_{3R}$.
The mass eigenmodes are explicitly given as
\bea
\left(
\begin{array}{c}
Q_3' \\
\Phi_3'
\end{array}
\right)_L &=&
\left(
\begin{array}{cc}
\cos \theta_L & \sin \theta_L \\
\sin \theta_L & \cos \theta_L
\end{array}
\right)
\left(
\begin{array}{c}
Q_3 \\
\Phi_3
\end{array}
\right)_L, \\
\left(
\begin{array}{c}
u_3' \\
\phi_3'
\end{array}
\right)_R &=&
\left(
\begin{array}{cc}
\cos \theta_R & \sin \theta_R \\
\sin \theta_R & \cos \theta_R
\end{array}
\right)
\left(
\begin{array}{c}
u_3 \\
\phi_3
\end{array}
\right)_R ,
\label{mixing}
\eea
where the mixing angles $\theta_L$ and $\theta_R$
are given by $\tan \theta_L = \omega/M_{\Phi}$ and
$\tan \theta_R = \bar{\omega}/M_{\phi}$
respectively.
Then the effective interaction Lagrangian is rewritten
in terms of these mass eigenmodes as
\bea
&-& {\cal L}_{\rm int} \nn \\
&=&
\lambda_* 
(\cos \theta_L \bar{\Phi}'_{3L} 
-\sin \theta_L \bar{Q}'_{3L} )
(\cos \theta_R \phi_{3R}' 
-\sin \theta_R u'_{3R} ) \tilde{H} \nn \\
& &
+ \lambda_* \bar{\Phi}_{3R} \phi_{3L} \tilde{H}  
+M_{\Phi}' \bar{\Phi}'_{3L} \Phi_{3R}
+M_{\phi}' \bar{\phi}_{3L} \phi_{3R}' \nn \\
& & + V(H),
\label{effint2}
\eea
where masses of the eigenmodes are given by
$M_{\Phi}' = M_{\Phi}/\cos \theta_L$
and
$M_{\phi}' = M_{\phi}/\cos \theta_R$.
The massless top quarks are identified with
$(Q'_{3L}, u'_{3R})$, and their effective 
Yukawa coupling with the Higgs field turns out to be
\be
Y_t^{\rm eff} \sim \lambda_* \sin \theta_L \sin \theta_R.
\ee
This coupling can be large enough, 
unless the mixing angles,
or the ratios $\omega/M$, are very small,
since the fixed point coupling $\lambda_*$ is
large in this model.

Thus the Higgs mass may be suppressed by the
conformal dynamics in a way compatible with the
large top Yukawa coupling.
However there are some issues to be concerned about
in the conformal Higgs model.
First one may wonder whether the 
$SU(3)_{\rm DSB}$ interaction affects the IR fixed point
and, moreover, may destroy the conformal invariance
fairly above the decoupling scale.
Indeed the conformal invariance is not exact, but
the beta function of the CFT gauge coupling is 
affected by the DSB gauge coupling $\alpha_{g'}$
at the two-loop level as
\be
\mu \frac{d \alpha_g}{d \mu} = \cdots + 4 \alpha_g^2 \alpha_g'.
\ee
Then the fixed point coupling is shifted
to $\alpha_g^* \sim (b_0 - 2 \alpha_g')/b_1$
effectively.
However the DSB sector may be regarded as 
an $N_c =3$ and $N_f = 6$ QCD and
$\alpha_{g'}$ becomes comparable
with $\alpha_g^*$ very near the dynamical scale 
$M_{\Phi}$.
Therefore the shift of the IR fixed point is very small.

We should also think about effects to the 
four-fermi couplings, 
since the DSB gauge interaction increases attractive
force among the fermions $\psi^A$ in the CFT sector.
In practice, the effective four-fermi interactions 
among the fermion $\psi$ are induced.
In the above model, however, 
the gauge coupling of the IR fixed point
is already close to the critical value $\alpha_{g,{\rm max}}$
without the  DSB interaction.
Therefore the RG flow of the four-fermi coupling 
enters the broken phase eventually, although the
running near the boundary is rather slow
\footnote{
Such a dynamical effect has been considered in 
the ``Postmodern Technicolor'' model \cite{ATW}, 
in which the QCD gauge interaction
drives the fixed point.
}.
However note that the DSB gauge interaction does not
affect the four-fermi interactions among $\Phi$
and $\phi$ directly.
The beta function for the four-fermi coupling $g_S$
(\ref{ladderRG}) is unchanged in the ladder or 
the large $N_c, N_f$  leading approximation.
Thus it is thought that the IR fixed point 
is not shifted remarkably
or destroyed through the DSB gauge interaction.

Next we also consider effects of the large anomalous
dimension on the scale of dynamically generated masses.
The four-fermi interactions given in the effective
Lagrangian (\ref{4fermi}) are also enhanced by the 
anomalous dimensions due to $SU(3)_{\rm CFT}$
interaction, since they include the fermions 
of the CFT sector,$\psi$, $\Phi$ and $\phi$.
This effect enlarges the scale of dynamical masses
\footnote{
The idea of enhancement of mass parameters by large 
anomalous dimensions have been developed in the walking 
Technicolor models \cite{walkingTC}.
}.
For example, the mass parameters $M_{\Phi}$ $(M_{\phi})$ and
$\omega$ $(\bar{\omega})$ are estimated in terms of the
dynamical scale $M_{\rm DSB}$ of the DSB interaction
as
\be
M_{\Phi} \sim \omega \sim \left(
\frac{M_{\rm DSB}}{\Lambda}
\right)^{2+\gamma_{\bar{\psi}\psi}^*}
M_{\rm DSB}.
\ee
Thus the extra matter fields in the CFT are decoupled at the mass
scale $M_{\Phi}$ and $M_{\phi}$, which are relatively larger
than the dynamical symmetry breaking scale $M_{\rm DSB}$.
Therefore the strong dynamics of the CFT is not responsible for 
the EW symmetry breaking, but solely reduces the radiative 
corrections to the Higgs mass parameter.

Provided that the effective Lagrangian also contains
the four-fermi coupling as
\be
\frac{\tilde{c}_{\Phi}}{\Lambda^2} \bar{\Phi}_{iA} \Phi^{iA}
(\bar{\psi}_A \cdot \psi^A),
\label{bad4fermi}
\ee
then the  fermion condensation 
$\langle \bar{\psi}_A \cdot \psi^A  \rangle$
generates the mass $\hat{M}_{\Phi}$ to the
fermion $\Phi$, which is enhanced as
\be
\hat{M}_{\Phi} \sim 
\left(
\frac{M_{\rm DSB}}{\Lambda}
\right)^{\gamma_{\bar{\psi}\psi}^*}
M_{\Phi}
\gg M_{\Phi}.
\ee
Then the decoupling mass scale is determined by
$\hat{M}_{\Phi}$ instead of $M_{\Phi}$.
Consequently the mixing angle is reduced to be 
$\omega/\hat{M}_{\Phi} \ll 1$ and, therefore,
the induced top Yukawa coupling turns out to be
too small.
This is the reason why we did not include
the interaction of (\ref{bad4fermi}) in the
effective Lagrangian (\ref{4fermi}).
We need to assume the coefficient $\hat{c}_{\Phi}$
of the four-fermi operator (\ref{bad4fermi}) 
to be suppressed with some reason.

\section{The Electro-Weak precision tests}

\subsection{Oblique corrections}

New interactions above the TeV scale, which is
required to push up the cutoff scale of the EW theory
to some higher scale, must be consistent with the
precision tests of the EW theory by the LEP experiments.
Especially the oblique corrections to the EW
gauge bosons put rather strong constraints to the
S and T parameters. 
In this subsection we shall evaluate these
parameters in the conformal Higgs model presented
in the previous section.

It has been known for some time that QCD-like gauge theories
induce an excessive correction to the S-parameter
\cite{PT,strongEW}.
Although it is a difficult dynamical problem to evaluate the oblique
corrections in general strong dynamics, there have been 
also some studies by using the DS equations.
According to the recent study \cite{HKY}, 
the S-parameter is decreased in the walking 
Technicolor models \cite{walkingTC} compared with
QCD-like models, however seems to be still large.
Thus models with dynamical EW symmetry breaking seem to have
a potential difficulty in satisfying the EW precision tests.
Contrary to this, neither the CFT interaction nor 
the DSB interaction does not induce the EW symmetry breaking
in the present scenario.
Therefore the huge correction to the S-parameter is not
generated through the strong dynamics.
This point is the essential difference from {\it e.g.}
the walking Technicolor \cite{walkingTC} and 
the conformal Technicolor \cite{LO}.

However the extra fermions $\Phi$ and $\phi$
couple with the Higgs field.
Moreover the weak isospin symmetry is largely broken,
since these fermions mix only with the top quark
fields.
Therefore a sizable correction to the $\rho$ parameter,
or the T-parameter, may be generated
through loop effect of these extra fermions.

\begin{widetext}
The oblique corrections are generated through
mixing among the EW doublet and the EW singlet
fields after the EWSB.
Explicitly, the VEV of the Higgs field,
$\langle H \rangle = (0 , v/\sqrt{2})$ leads 
the mass terms given by
\bea
& &
\left(
\begin{array}{ccc}
\bar{t}_L & \bar{\chi}'_{1L} & \bar{\phi}_L 
\end{array}
\right)
\left(
\begin{array}{ccc}
\lambda_* \sin \theta_L \sin \theta_R v/\sqrt{2} & 0 & 
- \lambda_* \sin \theta_L \cos \theta_R v/\sqrt{2} \\
- \lambda_* \cos \theta_L \sin \theta_R v/\sqrt{2} & M'_{\Phi} &
\lambda_* \cos \theta_L \cos \theta_R v/\sqrt{2} \\
0 & \lambda_* v/\sqrt{2} & M'_{\phi}
\end{array}
\right)
\left(
\begin{array}{c}
t_R \\
\chi_{1R} \\
\phi'_R
\end{array}
\right) \nn \\
& &
+ M'_{\Phi} \bar{\chi}_{2L} \chi_{2R}
+ \mbox{\it h.c.} ,
\label{massmatrix}
\eea
where the component fields are defined by
$Q'_{3L} = (t, b)_L$, $u'_{3R}=t_R$, 
$\Phi'_L = (\chi'_1, \chi'_2)_L$ and
$\Phi_R = (\chi_1, \chi_2)_R$.
Since the VEV for the EWSB $v$ is much smaller than 
$M_{\Phi}$ and $M_{\phi}$,
the top quark mass $m_t$ is given almost by
$\lambda_* \sin \theta_L \sin \theta_R v/\sqrt{2}
= Y_t^{\rm eff} v/\sqrt{2}$.
Therefore the mixing angles should satisfy
$\sin \theta_L \sin \theta_R 
\sim 1/\lambda_* $.
If we suppose $\bar{\omega} \sim \omega$
and $M_{\Phi} \sim M_{\phi} \sim M$ 
for simplicity,
the mixing angles in the explicit model 
should satisfy 
\be
\sin \theta_L  \sim \sin \theta_R \sim 1/\sqrt{\lambda_*}
\sim 0.6.
\label{theta}
\ee
\end{widetext}

Now mixing among the EW doublets and singlets is induced
through diagonalization of the mass matrix given by 
(\ref{massmatrix}).
The mixing decreases in proportion
to $v/M$ for a large $M$, as is seen from the mass matrix.
Therefore the oblique corrections are also 
vanishing for a sufficiently large $M$.
This is because the dynamically induced masses of $M$
for the extra matter fields are invariant under the
EW symmetry, and these fields just decouple from the
EW sector.
On the other hand, small but suitable amounts of 
oblique corrections should be induced so that the heavy
Higgs boson is compatible with the EW precision tests.
Thus the decoupling scale of the CFT sector
$M$ can be determined by this phenomenological 
consistency.
However, the scale $M$ should be relatively low 
in order to improve naturalness of the EW theory.
Therefore it is important to see whether the 
conformal Higgs model really allows the decoupling
scale less than a few TeV.

Now let us evaluate $\Delta T$
by calculating the one-loop corrections for
the self-energy of the EW gauge bosons.
In practice, the explicit model is very
similar to the top seesaw models \cite{topseesaw}
as far as the mixing mechanism is concerned, while
the top quark mass is not generated through the
seesaw mechanism.
So we may evaluate the oblique corrections in 
the same manner as done for the top seesaw models.
More explicitly, the oblique corrections induced by 
vectorlike fermions given in Ref.~\cite{loopcorr}
are also available.

Since the required corrections are small somehow,
we may consider only the cases with 
$\lambda_* v \ll M_{\Phi}, M_{\phi}$.
Then the mass terms given by (\ref{massmatrix}) 
can be approximated as 
\bea
& &m_t \bar{t}_L t_R 
+ M'_{\Phi} \bar{\chi}'_{1L} 
\left( \chi_{1R} 
- \frac{\lambda_* v}{\sqrt{2}M'_{\Phi}} 
\cos \theta_L \sin \theta_R t_R \right) \nn \\
& & + M'_{\phi}
\left( \bar{\phi}_L 
- \frac{\lambda_* v}{\sqrt{2}M'_{\phi}} 
\sin \theta_L \cos \theta_R \bar{t}_L \right)
\phi'_R.
\eea
It is seen that we need to take account of both mixings
between the left-handed fermions $(t_L, \phi_L)$
and between
the right-handed fermion $(t_R, \chi_{1R})$.

The contribution through the mixing of the
left-handed fermions is found to be
\be
\Delta T \sim \frac{3}{16 \pi^2 \alpha v^2}
\left\{
2 m_t^2 \ln \frac{{M'}^2_{\phi}}{m_t^2} + x_L^2
\right\}\left(\frac{x_L}{M'_{\phi}}\right)^2,
\ee
where 
$x_L = \lambda_* \sin \theta_L \cos \theta_R v/\sqrt{2}$.
In this expression, the fine structure constant 
$\alpha$ may be  evaluated as 
$\alpha^{-1}(M_Z) \sim 129$.
Similarly, the contribution through the mixing of the
right-handed fermions are found to be
\be
\Delta T \sim \frac{3}{16 \pi^2 \alpha v^2}
x_R^2 \left(\frac{x_R}{M'_{\Phi}}\right)^2,
\ee
where $x_R= \lambda_* \cos \theta_L \sin \theta_R v/\sqrt{2}$.
Thus both contributions by the heavy extra quarks
are positive.

If we take simply 
$M'_{\Phi} \simeq M'_{\phi} \simeq M$ and 
$x_L \simeq x_R \simeq x$,
the total correction of $\Delta T$ is
approximately given by
\be
\Delta T \sim \frac{3}{16 \pi^2 \alpha}
\left\{
\ln \frac{M^2}{m^2_t} + \frac{x^2}{m^2_t}
\right\}
\left( \frac{x}{M} \right)^2,
\label{deltaT}
\ee
where we used $2 m^2_t \simeq v^2$.
Here the parameter $x$ is not free but is related with
the fixed point Yukawa coupling $\lambda_*$ as
\be
x^2 \simeq m^2_t\left(
\frac{1}{\sin^2 \theta} - 1
\right) \simeq m^2_t (\lambda_* - 1).
\ee
If this oblique correction supplies 
$\Delta T = 0.25 \pm 0.1$ additionally 
to the SM correction, then the heavy Higgs
mass of $400 - 600$GeV becomes consistent
with the current evaluation of the EWPT
\cite{LEP,heavyhiggs}.
If we substitute $\lambda_* = 2.63$ into $\Delta T$
given by Eq.~(\ref{deltaT}), then the decoupling
mass scale may be determined to be
\be
1.5 \mbox{TeV} < M < 2.5 \mbox{TeV}.
\ee
It is noted that the consistent scale $M$ can appear 
at a relatively low energy region.
This is because that the fixed point Yukawa coupling
$\lambda_*$ is not very large in the explicit model.
Therefore, the amount of induced oblique correction
is model dependent.

Similarly, contribution to the S-parameter 
may be evaluated and is found to be
\be
\Delta S \leq \frac{1}{\pi}
\ln \frac{M^2}{m^2_t} \left(
\frac{x}{M}
\right)^2.
\label{deltaS}
\ee
This is much smaller than the contribution to 
$\Delta T$, since
\be
\frac{\Delta S}{\Delta T} \leq
\frac{16\pi}{3}\alpha \simeq  0.12.
\ee
This property is common with the top see-saw models
\cite{topseesaw}.
Thus the S-parameter is also consistent with
the EWPT. 
Here, we should say that these one-loop
analysis of the oblique correction is not so
definite indeed.
The higher order corrections by the QCD interaction
are not be negligible.
The CFT fermions also interact with the massive
gauge bosons rather strongly.
However such corrections are thought to be suppressed,
since the gauge boson mass is so heavy as the
decoupling scale $M$.
Therefore ambiguity of the one-loop analysis
would not be so large.
In any case, it seems that the present model with
the decoupling scale of a few TeV is viable.

\subsection{Z-boson decay width}

The EW precision tests also constrain the ratio
of decay widths of Z-boson,
$R_b \equiv \Gamma[Z \rightarrow b \bar{b}]/ 
\Gamma[Z \rightarrow \mbox{hadrons}]$
severely.
The deviation of coupling between the bottom quark 
and Z-boson $g_b$ from the SM value, $\delta g_b$,
is restricted roughly as $\delta g_b < 10^{-3}$.
Theoretically, this deviation can be also induced
through mixing between bottom quarks and the massive
extra fermions. 
Indeed there are such mixings in the explicit model
and $\delta R_b$ should be examined. 

So far we have ignored the original top and bottom 
Yukawa couplings, since these are suppressed and
not important to other effects. 
However we should add the bottom Yukawa term
$y_b \bar{Q}_{3L} b_R H$
to the effective Lagrangian given by (\ref{effint})
in order to see the mixing effect.
The original top and bottom Yukawa terms are
rewritten in terms of the mass eigenmodes into
\bea
-{\cal L}_{\rm Yukawa} 
&=&
y_t(\sin \theta_L \bar{\Phi}'_L + \cos \theta_L \bar{Q}'_{3L})
t_R \tilde{H} \nn \\
& & + 
y_b(\sin \theta_L \bar{\Phi}'_L + \cos \theta_L \bar{Q}'_{3L})
b_R H. 
\label{tbyukawa}
\eea
Therefore the EWSB induces the mixed mass terms given by
\be
\left(
\begin{array}{cc}
\bar{b}_L & \bar{\chi}'_{2L} 
\end{array}
\right)
\left(
\begin{array}{ccc}
y_b \cos \theta_L v/\sqrt{2} & 0 \\ 
y_b \sin \theta_L v/\sqrt{2} & M'_{\Phi} 
\end{array}
\right)
\left(
\begin{array}{c}
b_R \\
\chi_{2R} 
\end{array}
\right) 
+ \mbox{\it h.c.}.
\ee
Then it is shown that both of
$b_L$ and $b_R$ are mixed with components of the weak
doubles fermions $\Phi_L$ and $\Phi_R$ through
the EWSB.
Both of the $Zb_L\bar{b}_L$ and $Zb_R\bar{b}_R$ 
couplings induced the mixing effect are readily 
evaluated, since these are tree level contributions.
The deviations are found to be
\be
\delta g_{bL} \simeq  \delta g_{bR} \simeq
\left(
\frac{m_b}{M}
\right)^2
\leq 10^{-5},
\ee
which are sufficiently small.
Similarly, we may estimate the deviations in the
couplings of top quarks.
They are given roughly as
$\delta g_{tL} \simeq \delta g_{tR} \leq 10^{-2}$.

\subsection{Aspect of fine-tuning}

As is explained before this scenario leads to 
a relatively heavy Higgs. 
The mass is close to the triviality bound
corresponding with the scale $M$,
which are approximately $400-600$ GeV
Then the SM correction to the Higgs mass 
parameter becomes comparable with 
such a heavy mass itself,
when the scale $M$ of a few TeV acts as 
the cutoff scale.
Thus the fine-tuning due to the SM corrections
is not necessary.

Then, we should think about corrections
to the Higgs mass parameter by the CFT dynamics.
Indeed the Higgs mass is suppressed by the
anomalous dimension, however the correction is sizable
compared with the EW scale.
As was discussed in section II, the Higgs (mass)$^2$
at the decoupling scale $M$ is
given as Eq.~(\ref{scalarmass}).
This may be estimated as $O(1 - 10) \times M^2$,
which is too large to bring about the Higgs mass
of $400 - 600$ GeV.
Thus the initial mass parameter must be tuned somehow.
Since the degree of the fine-tuning \cite{BG}
is given roughly as
\be
\Delta = \frac{m_H^2(\Lambda)}{\tilde{m}_H^2(M)}
\frac{\delta \tilde{m}_H^2(M)}{\delta m_H^2(\Lambda)}
\sim \ln \left(
\frac{\Lambda^2}{M^2}\right)\frac{M^2}{\tilde{m}_H^2(M)},
\ee
it is found that fine-tuning of $O(1)$ \% is still
necessary in order to achieve the realistic EWSB
for the cutoff scale $\Lambda$ of, say, $O(100)TeV$.
Of course this cutoff scale cannot be taken extremely
high.
The model should be also replaced with some new 
physics, which probably does not contain the elementary
Higgs field.
However we do not discuss the ultraviolet
completion of the present model in this paper and
postpone it to future study.

\section{Conclusions and discussions}

In this paper we discussed a new scenario
in which the quadratic correction to the 
Higgs mass parameter is 
suppressed by a large anomalous dimension
endowed by interactions with a CFT sector.
With this mechanism, the cutoff scale can be
raised up sufficiently high so as to solve
the so-called little hierarchy problem of the SM. 
The strong dynamics of the CFT sector does 
not break the EW symmetry, but
radiative symmetry breaking of the Higgs field
as in the supersymmetric theories takes place.
The CFT sector just decouples from 
the SM sector by dynamical mass generation
at a few TeV scale.

It was shown explicitly that a large anomalous
dimension of the Higgs field can be realized
in a class of the gauge-Yukawa theories.
We analyzed the non-perturbative RG equations
to show it.
The quartic coupling of the Higgs fields is
rendered very irrelevant by the anomalous
dimension and converges to a fixed point value.
Due to strong dynamics of the CFT sector,
this fixed point coupling is rather large.
Therefore the scenario predicts a fairly
heavy mass for the SM Higgs,
which contrasts with the supersymmetric models
and the little Higgs models.

The anomalous dimension of the Higgs field
suppresses the Yukawa couplings in the SM as well.
Then compatibility with the large top quark mass 
becomes problematic.
We considered an explicit model, which we called 
``the conformal Higgs model'',
solving this problem due to mixing with 
top quark and the extra fermions of the
CFT sector.

The heavy Higgs boson is not consistent with the 
EWPT of the $(S, T)$ parameters, unless
a suitable amount of extra contribution is added.
However the mixing with the CFT fermions is
found to induce a proper oblique corrections
in the conformal Higgs model, if their
mass scale is given to be a few TeV.
Therefore the model predicts extra heavy
colored fermions and massive $SU(3)_C$ gauge 
bosons with a few TeV masses.
Correction to the decay width of Z-boson due to
the mixing effect is negligible.
Thus such extension of the SM also seems viable
phenomenologically.

The anomalous dimension of the Higgs field
may help to raise up the cutoff scale significantly.
However the model does not explain the scale of 
EWSB in general.
Naively the decoupling scale of a few TeV gives
also the Higgs mass scale.
Therefore the mass parameter must be tuned somewhat,
although the Higgs boson is fairly heavy.
So the aspect of naturalness is not very satisfactory,
and some improvement may be desired.
Meanwhile, the minimal supersymmetric SM also 
requires a similar degree of fine-tuning
\cite{lowsusy,fathiggs}.

Lastly we also mention further problems to be 
considered.
In the conformal Higgs model, the Higgs boson is
assumed to be point-like at least up to
some cutoff scale $\Lambda$.
However this scale cannot be taken extremely high,
and the ultra-violet completion of the model
should be considered.
One of the possible scenarios would be a composite
Higgs model.
It was also simply assumed that only top quark
is mixed with the extra fermions through the
Yukawa interactions.
It may be interesting to see whether the mixing effect
can be extended to other quarks/leptons than top quark.

It seems also interesting to give a five-dimensional
description of the conformal Higgs model, which is
suggested by the AdS/CFT correspondence \cite{AdS/CFT}.
The gauge structure of 
$SU(3)_{\rm CFT} \otimes SU(3)_C'$ and it's
spontaneous breaking to the diagonal subgroup
$SU(3)_C$ indicates a two-site deconstruction
of a five-dimensional model.
Indeed, the vector-like extra fermions may be
identified with the first 
Kaluza-Klein mode of the bulk top quark.
Moreover the massive gauge boson may be regarded
as the Kaluza-Klein mode of gluon in the bulk.
Study in this direction is now under way and the
results will be reported elsewhere. 

\section*{Acknowledgements}
The author is grateful to T.~Kobayashi, H.~Nakano, H.~Abe, 
M.~Tanabashi, K.~Yamawaki, M.~Kurachi, H.~D.~Kim, 
A.~Tsuchiya for valuable discussions and comments.
This work is supported in part by the Grants-in-Aid for 
Scientific Research (No.~14540256) and (No.~16028211).
from the Ministry of Education, Science, Sports and 
Culture, Japan.

\end{document}